\documentclass[a4paper,aps,prd,10pt,preprintnumbers,showkeys,twocolumn,superscriptaddress,nofootinbib,amsmath]{revtex4-1}
\usepackage{orcidlink,graphicx,cmap}
\usepackage[utf8]{inputenc}
\usepackage[T1]{fontenc}

\begin{document}

\title{Cesàro convergence of the high-order WKB method and its applications\\ to black-hole overtones and long-lived modes}

\author{Roman A.~Konoplya \orcidlink{0000-0003-1343-9584}}
\email{roman.konoplya@gmail.com}
\affiliation{Research Centre for Theoretical Physics and Astrophysics, Institute of Physics, Silesian University in Opava, Bezručovo náměstí 13, CZ-74601 Opava, Czech Republic}

\author{Jerzy Matyjasek \orcidlink{0000-0002-8312-592X}}
\email{jurek@kft.umcs.lublin.pl}
\affiliation{Institute of Physics, Maria Curie-Skłodowska University, pl. Marii Curie-Skłodowskiej 1, 20-031 Lublin, Poland}

\author{Alexander Zhidenko \orcidlink{0000-0001-6838-3309}}
\email{olexandr.zhydenko@ufabc.edu.br}
\affiliation{Research Centre for Theoretical Physics and Astrophysics, Institute of Physics, Silesian University in Opava, Bezručovo náměstí 13, CZ-74601 Opava, Czech Republic}
\affiliation{Centro de Matemática, Computação e Cognição (CMCC), Universidade Federal do ABC (UFABC), \\ Rua Abolição, CEP: 09210-180, Santo André, SP, Brazil}

\begin{abstract}
We develop a fully automatic Mathematica implementation of the black-hole WKB method at very high orders based on the Bender-Wu algorithm, which in principle is limited only by memory and computational time, and show that when pushed to sufficiently high order and improved by diagonal Padé approximants the method becomes efficient for two regimes which are usually regarded as difficult for the standard low-order WKB treatment: the first several overtones with $n>\ell$ and the very long-lived quasinormal modes of massive fields. At the same time, we show that this efficiency has a nontrivial limitation: for black-hole metrics belonging to the non-moderate class, especially when higher coefficients of the near-horizon parametrization become large, the WKB sequence may exhibit an apparent convergence to values which are nevertheless far from the accurate quasinormal frequencies. Thus, numerical stabilization of the WKB output alone is not always a sufficient criterion of correctness. However, we observe that although the WKB method with diagonal or near-diagonal Padé approximants does not exhibit monotonic convergence order by order, the corresponding Cesàro means become monotonically convergent once a sufficiently high WKB order is reached. This behavior may serve as an internal WKB criterion for the convergence of the method.
\end{abstract}

\maketitle

\section{Introduction}

The characteristic oscillation frequencies of black holes, known as quasinormal modes~\cite{Kokkotas:1999bd,Konoplya:2011qq}, have been studied extensively over the past decades because of their importance for black-hole physics and gravitational theory and, above all, because of the recent progress in gravitational-wave observations~\cite{LIGOScientific:2016aoc,LIGOScientific:2017vwq,LIGOScientific:2020zkf,EventHorizonTelescope:2019dse}. Even for the simplest Schwarzschild black hole, quasinormal modes cannot in general be calculated analytically, except in certain special limits or approximations~\cite{Bolokhov:2025rng}. One of the most efficient, reasonably accurate, and computationally inexpensive methods for calculating them is the WKB approach, first applied to the quasinormal-mode problem in~\cite{Schutz:1985km}. The method is based on a Taylor expansion of the wave function near the peak of the effective potential, matched to WKB expansions in the asymptotic regions. In its original first-order form, it yielded accurate results only in the high-multipole regime. It was soon extended to third order, leading to a considerable improvement in accuracy and already allowing a reasonable estimate of the fundamental mode ($\ell=2$, $n=0$) of the Schwarzschild and slowly rotating Kerr spacetimes~\cite{Iyer:1986np}. Much later, the method was extended to sixth order~\cite{Konoplya:2003ii}, which further improved its accuracy. Nevertheless, modes with larger damping rates, or with $n \geq \ell$, where $n$ is the overtone number and $\ell$ is the multipole number, have usually remained difficult to determine with satisfactory precision. The main advantage of the WKB method is its universality: it can be applied to essentially any effective potential with a single peak and monotonic decay at the asymptotic ends and therefore does not require substantial adaptation to each particular metric under consideration. In this sense, it is an automatic method. On the other hand, the WKB expansion is only asymptotic, so one is not guaranteed that each successive order will be more accurate than the previous one. Nevertheless, numerous studies~\cite{Bolokhov:2023bwm,Bolokhov:2023ruj,Bolokhov:2024ixe,Bolokhov:2025lnt,Bolokhov:2025egl,Bolokhov:2024bke,Bolokhov:2025fto,Skvortsova:2024wly,Skvortsova:2024atk,Lutfuoglu:2025hjy,Lutfuoglu:2025ljm,Lutfuoglu:2025ohb,Lutfuoglu:2025hwh,Lutfuoglu:2025qkt,Dubinsky:2025wns,Dubinsky:2025nxv,Dubinsky:2025fwv,Dubinsky:2024hmn} have shown that in the great majority of cases the WKB method, especially at sufficiently high order such as sixth order, is accurate enough for the numerical error to be much smaller than the physical effect under consideration. Thus, the attractiveness of a fast and automatic method has long competed with uncertainty in its numerical error. A major step forward was made in~\cite{Matyjasek:2017psv,Matyjasek:2019eeu}, where Padé approximants were introduced for the higher-order WKB formulas, significantly improving the accuracy and allowing even some modes with $\ell=n$ to be found reliably.

The next important step toward understanding the structure of very-high-order WKB corrections was taken by Hatsuda~\cite{Hatsuda:2019eoj}, who exploited the full analogy between the quasinormal-mode problem and the anharmonic-oscillator problem, first noted in~\cite{Zaslavsky:1991ug}, and applied Borel summation together with perturbative WKB expansion to reach very high WKB orders.

Given the current state of the WKB approach to quasinormal modes, our goal in this paper is to present an automatic Mathematica code that computes quasinormal frequencies by means of a very-high-order WKB technique. In principle, the method is not restricted to any fixed order; in practice, it is limited only by computational time and memory, together with the subsequent use of diagonal or near-diagonal Padé approximants.

Since the WKB method is usually already sufficiently efficient at relatively low orders for the least damped modes of massless fields, we focus here on two particularly important applications: higher overtones, especially those with $n\geq\ell$, and quasinormal modes of massive fields, both fundamental modes and overtones, in the regime where the damping rate approaches zero, giving rise to so-called quasi-resonances~\cite{Ohashi:2004wr,Konoplya:2004wg,Konoplya:2006br}. At first sight, the second problem may seem ill suited to the standard WKB approach proposed in~\cite{Galtsov:1991nwq}, because the effective potential may have either three turning points or only one, for instance when it develops an additional minimum or loses its peak altogether.

To address the first problem, namely the calculation of overtones, we formulate the perturbation problem in a parametrized background that cleanly separates weak-field information from genuine near-horizon deformations and then allow the symbolic code to generate WKB corrections to as high an order as needed. This parametrized approach is well motivated: the first several overtones are known to be highly sensitive to small static deformations near the event horizon~\cite{Konoplya:2022pbc}.

At the same time, the overtones play an important role in the early ringdown phase~\cite{Konoplya:2011qq}. In broad terms, while the fundamental mode is determined mainly by the geometry near the peak of the effective potential, the first few overtones, and hence the early ringdown, carry additional information about the near-horizon region, forming what one may call the ``sound of the event horizon''~\cite{Konoplya:2023hqb}.

Although the WKB method usually reproduces the photon-sphere mode rather well, it is much less accurate when used to describe modes dominated by near-horizon physics. For this reason, the black-hole parametrization developed in~\cite{Rezzolla:2014mua} provides a natural way to distinguish geometries for which WKB achieves a rapid effective convergence from those for which such convergence is difficult to obtain at intermediate orders.

The paper is organized as follows. Section~\ref{sec:parametrization} introduces the parametrized spherically symmetric black-hole geometry used throughout the paper. Section~\ref{sec:qnms} summarizes the quasinormal-mode problem and the methods employed here: Leaver's continued-fraction approach for accurate reference frequencies, the very-high-order WKB expansion with Padé resummation, and the relation to Borel and Borel--Padé summation. Section~\ref{sec:convergence} presents the numerical results and discusses the practical convergence properties of the high-order Padé sequence for overtones and massive quasi-resonant modes, including the distinction between true stabilization and apparent convergence. Finally, Section~\ref{sec:conclusions} contains the conclusions.

\section{Parametrized spherically symmetric black-hole geometry}\label{sec:parametrization}

Here, instead of restricting ourselves to any particular theory of gravity, we consider a general parametrization of asymptotically flat black holes in an arbitrary metric theory of gravity~\cite{Rezzolla:2014mua}. In this way, we can relate the convergence properties of the method to the geometry under consideration. This parametrization has already been used to approximate various numerical black-hole solutions~\cite{Konoplya:2019fpy,Kokkotas:2017ymc,Kokkotas:2017zwt} and extended to wormholes~\cite{Bronnikov:2021liv} higher-dimensional~\cite{Konoplya:2020kqb} and rotating~\cite{Younsi:2016azx} black holes. It is especially useful in the present context because it allows one to recast the metric functions, whether numerical or analytic but non-rational, into rational form~\cite{Konoplya:2022iyn,Konoplya:2023aph}. This is important because the coefficients of the resulting wave equation must also be rational in order to use the precise and convergent Frobenius method. We need the latter in order to compare the results obtained at different WKB orders with accurate quasinormal frequencies.

Following the general continued-fraction parametrization~\cite{Rezzolla:2014mua}, we write a static and spherically symmetric black-hole metric in the form
\begin{equation}
ds^2=-N^2(r)\,dt^2+\frac{B^2(r)}{N^2(r)}\,dr^2+r^2\left(d\theta^2+\sin^2\theta\,d\phi^2\right).
\end{equation}
The event-horizon radius corresponds to $r=r_0$, so that $N(r_0)=0$. It is convenient to compactify the radial coordinate according to
\begin{equation}
x=1-\frac{r_0}{r},
\qquad 0\leq x\leq 1,
\end{equation}
with $x=0$ at the event horizon and $x=1$ at spatial infinity. The lapse function is factorized as
\begin{equation}
N^2(x)=xA(x),
\qquad A(x)>0 \quad \text{for} \quad 0\leq x\leq 1,
\end{equation}
and the metric functions are expanded as
\begin{align}\nonumber
A(x)&=1-\epsilon(1-x)+(a_0-\epsilon)(1-x)^2+\widetilde A(x)(1-x)^3,\\
B(x)&=1+b_0(1-x)+\widetilde B(x)(1-x)^2.
\end{align}
The parameter
\begin{equation}\nonumber
\epsilon=\frac{2M-r_0}{r_0}
\end{equation}
measures the deviation of the horizon radius from the Schwarzschild value $2M$, while $a_0$ and $b_0$ encode the post-Newtonian asymptotics and are therefore tightly constrained by weak-field observations. The genuinely strong-gravity information is stored in the continued fractions
\begin{equation}
\widetilde A(x)=\frac{a_1}{1+\dfrac{a_2x}{1+\dfrac{a_3x}{1+\cdots}}},
\quad
\widetilde B(x)=\frac{b_1}{1+\dfrac{b_2x}{1+\dfrac{b_3x}{1+\cdots}}},
\end{equation}
whose coefficients dominate the near-horizon geometry. This split between asymptotic coefficients $(\epsilon,a_0,b_0)$ and near-horizon coefficients $(a_i,b_i)$ is precisely what makes the parametrization well suited for quasinormal-mode studies: it allows one to change the event-horizon geometry while keeping the weak-field region essentially fixed. Taking into account modern post-Newtonian constraints we usually imply that $a_0 \approx 0$ and $b_0\approx0$.

For the purposes of the present paper it is useful to distinguish three reference classes of black holes~\cite{Konoplya:2020hyk,Konoplya:2022pbc}.

\paragraph*{1. Schwarzschild black holes.}
This is the undeformed reference point obtained by setting all deformation parameters to zero,
\begin{equation}
\epsilon=a_0=b_0=a_1=b_1=0.
\end{equation}

\paragraph*{2. Moderately deformed black holes.}
These are parametrized black holes for which the metric functions vary relatively slowly in the near-horizon zone. In such cases the deformation of the effective potential is smooth and the quasinormal spectrum changes perturbatively: the fundamental mode and the first overtones change approximately at the same rate and there is no dramatic enhancement of overtone sensitivity~\cite{Konoplya:2022tvv}. This is the regime in which standard low-order WKB formulas are expected to behave in the most familiar way. All the parametriation coefficients are usually very small in this case.

\paragraph*{3. Nonmoderately deformed black holes.}
These geometries are close to Schwarzschild in the weak-field region and may satisfy the same size-to-mass relation and post-Newtonian constraints, so that one keeps $\epsilon$, $a_0$, and $b_0$ zero or close to zero, $a_1$, and $b_1$ very small, but one or more higher near-horizon coefficients (usually $a_2$, $b_2$ and/or higher) in the continued fractions are large enough to deform the geometry sharply in a narrow region close to $r=r_0$. This is the most interesting class. It isolates genuinely near-horizon physics: the fundamental mode can remain almost unchanged, whereas the first few overtones may shift drastically, even when the difference between the corresponding effective potentials is tiny outside the immediate vicinity of the horizon. In this sense such backgrounds behave as Schwarzschild mimickers for the fundamental mode and optical observables, but not for the overtone spectrum.

The adjective ``nonmoderate'' will be used below in a broader sense for any background with a rapid near-horizon variation, but the asymptotically Schwarzschild-like subclass ($\epsilon=a_0=b_0=0$) is the cleanest arena for our analysis because it removes the trivial possibility that large spectral changes are caused simply by a large asymptotic deformation.

\section{Quasinormal modes}\label{sec:qnms}

\subsection{Precise quasinormal frequencies via the Leaver method}

To assess the accuracy of the very-high-order WKB expansion we need reference frequencies whose numerical uncertainty is negligible on the scale of the effects discussed below. For this purpose we use Leaver's continued-fraction method~\cite{Leaver:1985ax}, which is particularly convenient for the parametrized black-hole backgrounds~\cite{Konoplya:2022iyn,Konoplya:2022tvv,Bolokhov:2023ozp} introduced in the previous section. In this case the coefficients of the perturbation equation become rational functions of $x=1-r_0/r$, and the Frobenius construction can be implemented in a systematic and stable way~\cite{Konoplya:2011qq}. In fact, Leaver's method allows one to find quasinormal modes with as high accuracy as necessary, because it is based on a convergent procedure.

For a massive scalar field, and frequently for fields of other spin, the radial function satisfies
\begin{equation}
\frac{d^2\Psi}{dr_*^2}+\left(\omega^2-V(r)\right)\Psi=0,
\label{eq:radial-leaver}
\end{equation}
and the quasinormal-mode boundary conditions take the form
\begin{equation}
\Psi\propto
\begin{cases}
 e^{-i\omega r_*}, & r_*\to -\infty \quad (r\to r_0),\\
 e^{+i\chi r_*}, & r_*\to +\infty \quad (r\to \infty),
\end{cases}
\quad
\chi=\sqrt{\omega^2-\mu^2}.
\label{eq:bc-massive-leaver}
\end{equation}
Thus the solution is purely ingoing at the event horizon and purely outgoing at spatial infinity. For massive fields the choice of the square-root branch matters: In order to have the outgoing wave at infinity, $\mathrm{Re}(\chi)$ and $\mathrm{Re}(\omega)$ must have the same sign. In this way
we select the branch that continuously connects to the massless limit, so that $\chi\to\omega$ when $\mu\to0$. This prescription is essential for quasi-resonant modes, for which $|\mathrm{Im}(\omega)|$ may become very small and the asymptotic discrimination between outgoing and nonphysical solutions becomes numerically delicate.

Leaver's method incorporates conditions (\ref{eq:bc-massive-leaver}) directly into the ansatz. Using the compact coordinate introduced earlier, we write
\begin{equation}
\Psi=H(x)\sum_{n=0}^{\infty} d_n x^n,
\qquad x=1-\frac{r_0}{r},
\label{eq:frob-ansatz}
\end{equation}
where the prefactor is chosen so that $H\propto e^{-i\omega r_*}$ near $x=0$ ($r=r_0$) and $H\propto e^{+i\chi r_*}$ as $x\to1$ ($r\to\infty$). The remaining series is regular at the horizon and, for the correct quasinormal frequencies, defines the minimal solution in the sense of the continued-fraction construction.

Substituting Eq.~(\ref{eq:frob-ansatz}) into Eq.~(\ref{eq:radial-leaver}) yields a linear recurrence relation for the Frobenius coefficients. In generic parametrized metrics this relation need not be three-term from the beginning; rather, one typically obtains
\begin{equation}
\sum_{j=0}^{p} c_{j,n}d_{n-j}=0,
\qquad n\ge p-1,
\label{eq:general-recurrence}
\end{equation}
with $p>2$, because the rational structure of the metric functions generates several neighboring couplings between the coefficients. At the same time, the Frobenius series centered at the horizon converges only up to the nearest singular point of the radial equation. Therefore, a single series representation reaches spatial infinity only when there are no additional singular points between the event horizon and infinity. In the general case one must analytically continue the solution by re-expanding it at a sequence of regular intermediate points and matching the neighboring series, following the midpoint procedure used in continued-fraction calculations~\cite{Rostworowski:2006bp}. One then reduces the higher-order recurrence by Gaussian elimination to the standard three-term form
\begin{eqnarray}\label{eq:three-term-recurrence}
\alpha_0 d_1+\beta_0 d_0&=&0,
\\\nonumber
\alpha_n d_{n+1}+\beta_n d_n+\gamma_n d_{n-1}&=&0,
\qquad n\ge 1,
\end{eqnarray}
where $\alpha_n$, $\beta_n$, and $\gamma_n$ depend on the frequency $\omega$, the field mass $\mu$, the multipole number $\ell$, and the black-hole parameters entering the continued-fraction expansion of the background geometry.

The quasinormal spectrum can then be found from the infinite continued-fraction equation
\begin{equation}
0=\beta_0-\frac{\alpha_0\gamma_1}{\beta_1-\dfrac{\alpha_1\gamma_2}{\beta_2-\dfrac{\alpha_2\gamma_3}{\beta_3-\cdots}}}.
\label{eq:leaver-cf}
\end{equation}
Its roots in the complex $\omega$ plane give the desired quasinormal frequencies. In practice this method is far more robust than direct shooting for long-lived massive modes, because one does not have to resolve a tiny admixture of the unwanted asymptotic branch numerically; the analytic structure of the local solutions already suppresses it.

For the quasi-resonant regime and for higher overtones, however, the raw continued fraction may converge rather slowly. In order to improve its convergence, we use an asymptotic estimate for the tail of the continued fraction, which substantially accelerates convergence for weakly damped or highly excited modes~\cite{Nollert:1993zz,Zhidenko:2006rs}. In this form the Frobenius--Leaver method provides the high-precision benchmark data used throughout the paper to decide whether the WKB and Padé sequences converge to the correct physical answer or merely stabilize numerically.

\subsection{High-order WKB expansion and Padé resummation}

It is useful to regard the WKB method not merely as a low-order formula, but rather as a high-order asymptotic construction supplemented by a rational resummation procedure.
In what follows, we show that, when developed to sufficiently high order, the WKB method can yield accurate quasinormal frequencies for a broad class of black-hole geometries, not only for massless fields but also for massive ones, provided that the effective potential retains a local maximum.

\begin{figure}
\resizebox{\linewidth}{!}{
\begin{tikzpicture}[x=1.25cm,y=1.15cm]
\draw[->,black,thick] (-3.5,0) -- (3.7,0) node[right] {$r_*$};
\draw[->,black,thick] (0,-1.0) -- (0,1.15) node[above] {$Q(r_*)$};
\draw[black,thick,smooth] plot coordinates {
(-3.3,0.78) (-2.5,0.78) (-1.7,0.45) (-0.92,0) (0,-0.62) (0.92,0) (1.7,0.35) (2.5,0.55) (3.3,0.55)
};
\draw[black,densely dashed] (-0.92,0) -- (-0.92,-0.62);
\draw[black,densely dashed] (0.92,0) -- (0.92,-0.62);
\draw[black,densely dashed] (0,-0.62) -- (0,0);
\fill[black] (-0.92,0) circle (1.2pt);
\fill[black] (0.92,0) circle (1.2pt);
\fill[black] (0,-0.62) circle (1.2pt);
\draw[black,densely dotted] (-3.3,0.78) -- (-2.35,0.78);
\draw[black,densely dotted] (2.35,0.55) -- (3.3,0.55);
\node[black,below] at (-0.92,-0.08) {$r_{*1}$};
\node[black,below] at (0.92,-0.08) {$r_{*2}$};
\node[black,below right] at (0,-0.62) {$r_{*0}$};
\node[black,above left] at (-2.55,0.78) {$Q\to\omega^2$};
\node[black,above right] at (2.1,0.55) {$Q\to\chi^2$};
\end{tikzpicture}
}%
\caption{A schematic profile of $Q(r_*)=\omega^2-V(r_*)$, corresponding to a single-barrier effective potential $V(r_*)$. The local minimum of $Q$ is located at $r_{*0}$, and the turning points $r_{*1}$ and $r_{*2}$ are defined by $Q=0$. Near the horizon $Q\to\omega^2$, while for the massive-field case at spatial infinity $Q\to\chi^2=\omega^2-\mu^2$; in the massless limit $\chi=\omega$.}
\label{fig:wkb-barrier}
\end{figure}
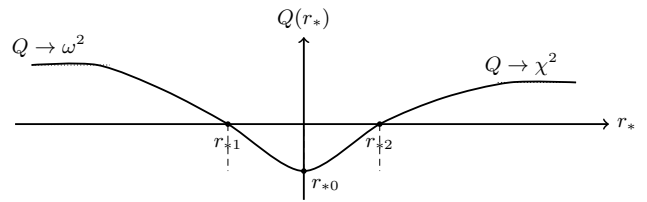

For the class of problems to which the standard black-hole WKB approach applies~\cite{Konoplya:2003ii,Matyjasek:2019eeu,Konoplya:2019hlu}, the radial equation can be written as
\begin{equation}
\frac{d^2\Psi}{dr_*^2}+Q(r_*)\Psi=0,
\qquad
Q(r_*)=\omega^2-V(r_*),
\label{eq:wkb-barrier-q}
\end{equation}
where the function $Q(r_*)$ has a single local minimum at $r_*=r_{*0}$; equivalently, the effective potential $V(r_*)$ has a single local maximum there. In our asymptotically flat problem the left asymptotic region corresponds to the vicinity of the event horizon, where $V\to0$ and therefore $Q(r_*)\to\omega^2$ as $r_*\to-\infty$. For the massive-field case considered below, the right asymptotic region corresponds to spatial infinity, where $V\to\mu^2$ and therefore $Q(r_*)\to\chi^2=\omega^2-\mu^2$ as $r_*\to+\infty$; in the massless limit one has $\chi\to\omega$. A schematic profile of $Q(r_*)$ of the type underlying the WKB construction is shown in Fig.~\ref{fig:wkb-barrier}.

Matching the WKB solutions across the turning points and analytically continuing the scattering problem to the quasinormal-mode sector leads to the quantization rule
\begin{equation}
\frac{iQ_0}{\sqrt{2Q_0''}}-\sum_{j=2}^{N}\Lambda_j=\mathcal K,
\qquad
\mathcal K\equiv n+\frac{1}{2},
\label{eq:wkb-master}
\end{equation}
where $Q_0=Q(r_{*0})$, the prime denotes differentiation with respect to $r_*$, and each correction $\Lambda_j$ is an algebraic expression built from derivatives of $Q$ at the peak up to order $2j$. Since $Q(r_*)=\omega^2-V(r_*)$, one may rewrite Eq.~(\ref{eq:wkb-master}) in the explicit form
\begin{equation}
\omega^2=V_0-i\mathcal K\sqrt{-2V_0''}-i\sqrt{-2V_0''}\sum_{j=2}^{N}\Lambda_j,
\label{eq:wkb-omega-explicit}
\end{equation}
where $V_0=V(r_{*0})$. Thus, once the derivatives of the effective potential at $r_{*0}$ are known, the problem reduces to a purely symbolic construction of the higher-order corrections. In particular, an $N$th-order approximation requires derivatives of the potential up to order $2N$.

The above WKB formulation can be mapped to the perturbative eigenvalue problem for a quantum anharmonic oscillator. In this representation, the quasinormal-mode condition is translated into an energy quantization problem, while the WKB corrections $\Lambda_j$ become the perturbative coefficients of the corresponding oscillator problem. The relation between black-hole quasinormal modes and the anharmonic oscillator was and the corresponding pointed out in~\cite{Zaslavsky:1991ug,Bonatsos:1992he} and later exploited in the high-order WKB analysis of~\cite{Matyjasek:2019eeu,Hatsuda:2019eoj}.

The practical advantage of this reformulation is that one can use the Bender--Wu algorithm~\cite{Bender:1973rz}, originally developed for high-order perturbation theory of anharmonic oscillators, to generate the coefficients recursively to very large orders. Instead of deriving each higher-order WKB term separately by hand, one solves an algebraic recursion problem for the local wave function and the corresponding perturbative eigenvalue, and then translates the resulting series back into the WKB expansion for $\omega^2$. In practice, this makes a fully automatic very-high-order implementation feasible through the Wolfram Mathematica{\textregistered} package~\cite{Sulejmanpasic:2016fwr}, which computes the higher-order WKB corrections recursively.

For practical calculations it is advantageous to regard the right-hand side of Eq.~\eqref{eq:wkb-omega-explicit} as a formal truncated series in an auxiliary bookkeeping parameter $\lambda$,
\begin{eqnarray}\nonumber
P_N(\lambda)&=&V_0-i\mathcal K\sqrt{-2V_0''}\,\lambda
-i\sqrt{-2V_0''}\sum_{j=2}^{N}\lambda^j\Lambda_j,
\\
\omega^2&\approx& P_N(1).
\label{eq:wkb-pn}
\end{eqnarray}
The direct value $P_N(1)$ is often useful, but for higher overtones or slowly damped massive modes it may oscillate with order and therefore provide an unreliable picture of convergence. For this reason we adopt the Padé reformulation and replace the polynomial by a rational approximant
\begin{equation}
\mathcal P^{\tilde m}_{\tilde n}(\lambda)=
\frac{\sum_{j=0}^{\tilde m} A_j\lambda^j}{1+\sum_{j=1}^{\tilde n} B_j\lambda^j},
\qquad
\tilde m+\tilde n=N,
\label{eq:wkb-pade}
\end{equation}
whose Taylor expansion agrees with $P_N(\lambda)$ through order $N$. The quasinormal frequency is then approximated by $\omega^2\approx \mathcal P^{\tilde m}_{\tilde n}(1)$. In practice the diagonal and near-diagonal approximants are the most informative, because their clustering gives a much better indication of stabilization than the raw order-by-order WKB sequence.

\subsection{Borel and Borel--Padé summation}

The discussion of the previous subsection suggests a natural further step. Once the WKB output is viewed as a formal series in the auxiliary parameter $\lambda$,
\begin{equation}
P(\lambda)=\sum_{k=0}^{\infty} c_k\lambda^k,
\label{eq:wkb-formal-series}
\end{equation}
with $P_N(\lambda)$ representing its truncation at order $N$, one may ask whether this large-order series admits a well-defined resummation beyond ordinary Padé improvement. Hatsuda's analysis~\cite{Hatsuda:2019eoj} addresses precisely this question by treating the WKB expansion as a genuinely asymptotic object and relating the black-hole quasinormal-mode problem to the perturbation theory of an anharmonic oscillator. In the standard single-barrier cases, such as the Schwarzschild and Reissner--Nordström backgrounds, this program shows that the high-order WKB data can indeed be resummed to the correct quasinormal frequencies,
\begin{equation}
\omega^2=P(1).
\end{equation}

The corresponding Borel transform is defined by~\cite{Hatsuda:2019eoj}
\begin{equation}
\mathcal B[P](\zeta)=\sum_{k=0}^{\infty}\frac{c_k}{k!}\zeta^k,
\label{eq:borel-transform}
\end{equation}
and, whenever this transform can be analytically continued along the positive real axis without obstructing singularities, the resummed value is recovered from the Laplace-type integral
\begin{equation}
\mathcal S_{\mathrm{Borel}}[P](\lambda)=\int_{0}^{\infty} e^{-t}\mathcal B[P](\lambda t)\,dt.
\label{eq:borel-sum}
\end{equation}
At the physical point $\lambda=1$, the quantity $\mathcal S_{\mathrm{Borel}}[P](1)$ provides the Borel-resummed estimate for $\omega^2$,
\begin{equation}
\omega^2=\mathcal S_{\mathrm{Borel}}[P](1).
\end{equation}

In practice the analytic continuation of $\mathcal B[P](\zeta)$ is usually not known in closed form, so one replaces the truncated Borel transform by a Padé approximant and then evaluates the integral in Eq.~\eqref{eq:borel-sum}. This leads to a Borel--Padé resummation. From the point of view of the present paper, this should be regarded as a refinement of the Padé strategy discussed above rather than as a separate framework: ordinary Padé resummation acts directly on $P_N(\lambda)$, whereas Borel--Padé first reorganizes the same perturbative information into a form more suitable for large-order continuation.

Conceptually, this observation is important for our purposes. It means that the very-high-order WKB coefficients contain more information than is visible in a plain truncation, and that a slow or irregular order-by-order (that is, non-monotonic via orders) convergence does not by itself imply the absence of a meaningful resummed answer. In the applications considered below, diagonal and near-diagonal Padé approximants remain the primary diagnostic tool, while Borel or Borel--Padé summation provides a natural next step whenever the Padé family stabilizes too slowly to yield a decisive conclusion.

\section{Quasinormal modes and practical convergence of the WKB results at different orders}\label{sec:convergence}

\begin{table}
\resizebox{\linewidth}{!}{
\begin{tabular}{lcc}
 \hline
 \hline
$\mu$ & accurate (Frobenius) & 500 order WKB\\
\hline
 0.00 & 0.5858722665-0.1953199778 i & 0.5858722665-0.1953199778 i \\
 0.10 & 0.5881086314-0.1939759577 i & 0.5881086314-0.1939759577 i \\
 0.20 & 0.5948313225-0.1899141472 i & 0.5948313225-0.1899141472 i \\
 0.30 & 0.6060798071-0.1830411334 i & 0.6060798071-0.1830411334 i \\
 0.40 & 0.6219138168-0.1731865712 i & 0.6219138168-0.1731865712 i \\
 0.50 & 0.6423970327-0.1600792413 i & 0.6423970327-0.1600792413 i \\
 0.60 & 0.6675543875-0.1433154198 i & 0.6675543875-0.1433154198 i \\
 0.70 & 0.6972793640-0.1223477650 i & 0.6972793640-0.1223477650 i \\
 0.80 & 0.7312129421-0.0965702672 i & 0.7312129421-0.0965702672 i \\
 0.90 & 0.7687311615-0.0655074211 i & 0.7687312953-0.0655027670 i \\
 0.91 & 0.7726514853-0.0620995889 i & 0.7726427056-0.0620829489 i \\
 0.92 & 0.7765993717-0.0586363997 i & 0.7764448484-0.0586872822 i \\
 0.93 & 0.7805740601-0.0551178038 i & 0.7899202536-0.0654764417 i \\
 0.94 & 0.7845747955-0.0515437756 i & \\
 0.95 & 0.7886008298-0.0479143126 i & \\
 0.96 & 0.7926514226-0.0442294345 i & \\
 0.97 & 0.7967258421-0.0404891824 i & \\
 0.98 & 0.8008233659-0.0366936179 i & \\
 0.99 & 0.8049432820-0.0328428224 i & \\
 1.00 & 0.8090848888-0.0289368960 i & \\
 1.01 & 0.8132474965-0.0249759569 i & \\
 1.02 & 0.8174304272-0.0209601401 i & \\
 1.03 & 0.8216330152-0.0168895967 i & \\
 1.04 & 0.8258546079-0.0127644930 i & \\
 1.05 & 0.8300945649-0.0085850103 i & \\
 1.06 & 0.8343522607-0.0043513357 i & \\
 1.07 & 0.8386270829-0.0000636807 i & \\
 \hline
 \hline
\end{tabular}
}
\caption{Schwarzschild black hole ($r_0=1$): Quasinormal frequencies of the massive scalar field ($\ell=1$, $n=0$) calculated using the accurate Frobenius method vs 500 order WKB with the diagonal Padé approximant. For $\mu=0.93$, 500 order WKB is not sufficient to reproduce correctly the second decimal place due to slow convergence, however 1500 order gives the following approximation $\omega_0=0.7794978761 - 0.0553404140i$, reproducing correctly three decimal places. For $\mu \gtrsim 0.94$ the effective potential is monotonous and does not have any local maxima.}\label{table:compare}
\end{table}

The automatic Mathematica code used in this section has three levels. First, it generates numerically the derivatives with respect to the tortoise coordinate of the effective potential at the peak. Second, it constructs the WKB corrections up to a chosen order. Third, it builds the corresponding Padé approximants for $\omega^2$. This is the version of the method relevant for the data below, because in the regimes of interest (overtones with $n>\ell$ and long-lived massive modes) the raw WKB truncation is usually much less informative than the diagonal or near-diagonal Padé sequence~\cite{Matyjasek:2019eeu,Hatsuda:2019eoj,Konoplya:2004wg}.

For the numerical discussion we denote by $\omega_N$ the quasinormal frequency obtained from the diagonal or (sub)diagonal Padé approximant constructed from the $N$th-order WKB expansion,
\begin{equation}
\omega_N^2=\mathcal P^{\tilde m}_{\tilde n}(1),
\qquad
\tilde m = [N/2],\quad \tilde n=N-\tilde m
\label{eq:wkb-pade-sequence}
\end{equation}
where $[N/2]$ is the integer part of the order divided by two, so that for even $N=2k$ $\tilde m=\tilde n=k$ and for odd $N=2k+1$ $\tilde m=\tilde n-1=k$.

The data below show that the sequence $\{\omega_N\}$ usually does not converge monotonically with $N$. A more accurate description is non-monotonic convergence of the Padé sequence, often accompanied by stabilization of its local averages. This behavior is specific to diagonal and near-diagonal Padé approximants; for the raw WKB truncations $P_N(1)$, merely increasing the WKB order does not lead to comparable stabilization~\cite{Matyjasek:2019eeu,Hatsuda:2019eoj}.

To describe the slow drift visible in the plots, it is useful to introduce the running arithmetic mean
\begin{equation}
\overline{\omega}_{N,\Delta}=\frac{1}{\Delta+1}\sum_{s=0}^{\Delta}\omega_{N-s},
\label{eq:wkb-running-mean}
\end{equation}
where $\Delta$ is chosen large enough to smooth the short-scale oscillations but small enough to preserve the slow drift of the central value. The special case
\begin{equation}
\sigma_N=\overline{\omega}_{N,N}=\frac{1}{N+1}\sum_{k=0}^{N}\omega_k
\label{eq:wkb-cesaro}
\end{equation}
is the Cesàro mean. If $\sigma_N\to\sigma$, the sequence is said to be Cesàro summable to $\sigma$~\cite{HardyDivergent}.

\begin{figure*}
\resizebox{\linewidth}{!}{\includegraphics{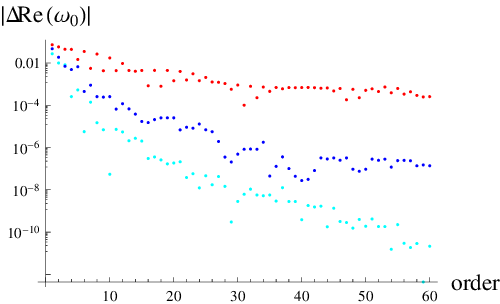}\includegraphics{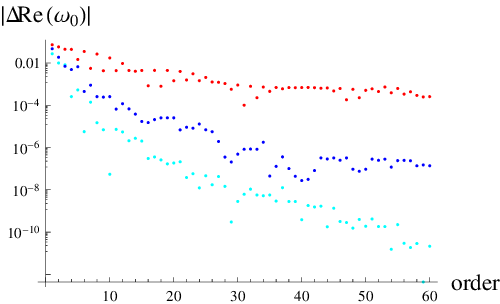}}
\resizebox{\linewidth}{!}{\includegraphics{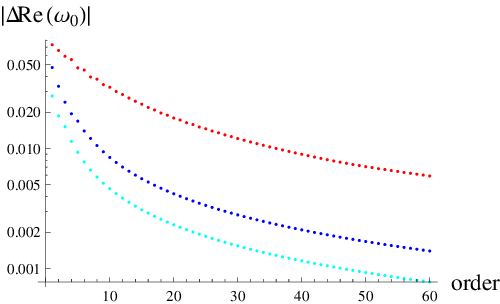}\includegraphics{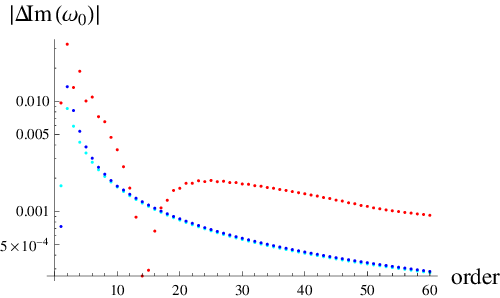}}
\caption{Schwarzschild black hole ($r_0=1$): error in real and imaginary part of the quasinormal frequency of the massive scalar field ($\ell=1$, $n=0$) for the diagonal (even orders) and subdiagonal (odd orders) Padé approximants of different orders (upper panels) and for the Cesàro mean (lower panels). From bottom to top we have $\mu=0.7$ (cyan), $\mu=0.8$ (blue), $\mu=0.9$ (red). Although the error does not decrease monotonically with the WKB order, it generally becomes smaller at higher orders for all values of $\mu$, until the local maximum disappears. In the vicinity of this point, the convergence becomes slower. Nevertheless, sufficiently high WKB orders still provide a good approximation.}\label{fig:convergencemu}
\end{figure*}

In Fig.~\ref{fig:convergencemu} we plot the absolute errors in the real and imaginary parts of the quasinormal frequency of the massive scalar field at each WKB order, using the accurate Frobenius results as the benchmark. The plots show that the Cesàro mean does indeed stabilize the sequence $\{\omega_N\}$. This does not imply monotonic convergence at every order: For example, the case $\mu r_0=0.9$ still exhibits non-monotonic behavior of the absolute error. However, at sufficiently high WKB order the overall trend is a steady decrease of the error. The drawback is that the Cesàro means converge much more slowly than the appropriate diagonal or near-diagonal Padé approximants. Therefore, whenever the mean values do converge, the corresponding Padé-improved high-order WKB results provide more accurate estimates than the Cesàro averages themselves.

\begin{figure*}
\resizebox{\linewidth}{!}{\includegraphics{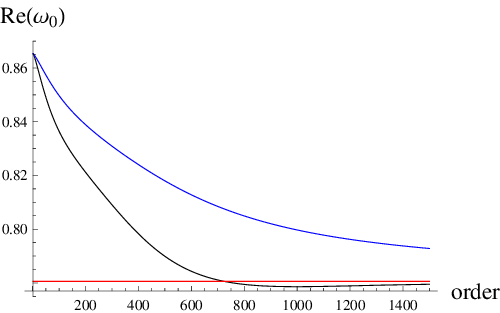}\includegraphics{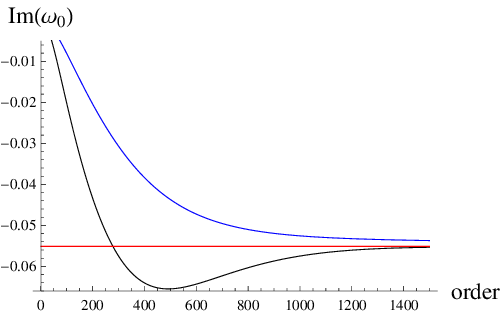}}
\caption{Schwarzschild black hole ($r_0=1$): error in real and imaginary part of the quasinormal frequency ($\ell=1$, $n=0$) of the massive scalar field ($\mu=0.93$) for the diagonal Padé approximants of different orders (black) and the corresponding Cesàro means (blue, top) vs accurate value (red horizontal line).}\label{fig:convergencemu093}
\end{figure*}

Another useful way to visualize the convergence is to restrict attention to the diagonal Padé subsequence, which is naturally defined only for even WKB orders. In this way one suppresses part of the irregular order-to-order scatter and obtains a clearer picture of the large-order behavior. The resulting sequence is still not strictly monotonic, but its oscillations are noticeably smoother, and its deviation from the accurate Frobenius value is generally smaller than that of the corresponding Cesàro means. Figure~\ref{fig:convergencemu093} shows this comparison for a particularly difficult case in which the convergence is extremely slow. There the diagonal Padé approximants already reveal the correct trend at finite order, whereas the Cesàro averages provide a more conservative but slower route to stabilization. Thus the Cesàro mean is useful mainly as an auxiliary diagnostic of convergence, while the diagonal Padé subsequence usually gives the more accurate finite-order estimate of the quasinormal frequency.

\begin{figure}
\resizebox{\linewidth}{!}{\includegraphics{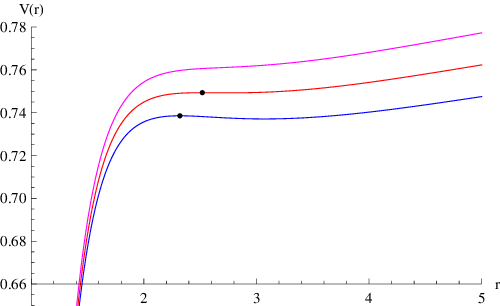}}
\caption{Schwarzschild black hole ($r_0=1$): Effective potential for the massive scalar field ($\ell=1$) for $\mu=0.92$ (blue), $\mu=0.93$ (red), $\mu=0.94$ (magenta). The diagonal Padé approximants show convergence even when the potential has only a local maximum (black dots).}\label{fig:compare}
\end{figure}

The convergence of the high-order Padé approximants reveals another remarkable feature. Strictly speaking, the standard black-hole WKB derivation is formulated for two turning points, whereas the effective potential of a massive field may develop three turning points. Nevertheless, as long as the potential still possesses a local maximum (see Fig.~\ref{fig:compare}), the diagonal Padé sequence often yields accurate quasinormal frequencies (see Table~\ref{table:compare}), even when the effective potential no longer has the standard isolated single-barrier shape. A plausible interpretation is that the high-order derivatives evaluated at the point of maximum continue to capture the leading local scattering data, while the additional turning point alters the global matching only through corrections that are partly encoded by the Padé resummation. At present we view this as an empirical property of the resummed high-order WKB scheme rather than as a rigorous derivation of a three-turning-point formula~\cite{Iyer:1986np,Konoplya:2003ii,Hatsuda:2019eoj}.

\begin{figure*}
\resizebox{\linewidth}{!}{\includegraphics{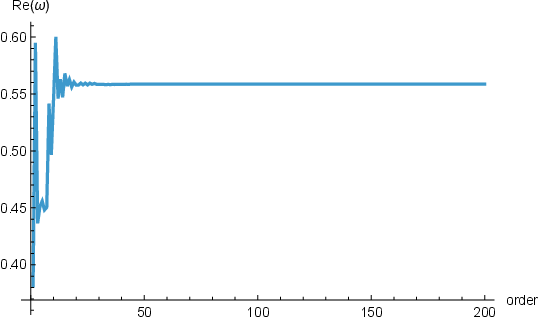}\includegraphics{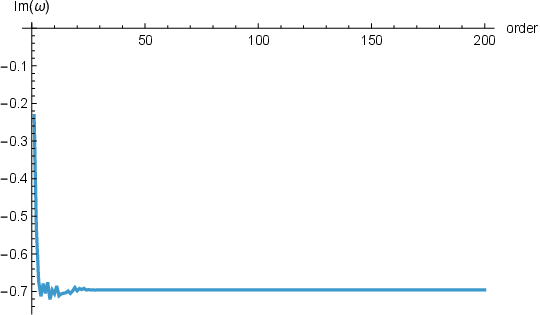}}
\caption{Example of rapid convergence to the accurate value: $r_0=1$, $\epsilon=0$, $a_{1}=0.5$, $a_{2}=2$ ($\ell=n=1$). The Frobenius method yields the quasinormal frequency $\omega = 0.55863 - 0.695763 i$, which is already reproduced by the 200th-order WKB approximation, $\omega = 0.558630 - 0.695763 i$.}\label{fig:A2}
\end{figure*}

Whenever genuine convergence is present, it can usually be identified not only in the diagonal Padé subsequence associated with even WKB orders, but also in the neighboring near-diagonal approximants constructed from odd orders. Although the two subsequences may differ noticeably at intermediate orders, for sufficiently large truncation order they tend to approach the same limiting frequency. As shown in Fig.~\ref{fig:A2}, in this regime both even- and odd-order Padé approximants fluctuate around the accurate Frobenius value with progressively decreasing amplitude.

\begin{figure*}
\resizebox{\linewidth}{!}{\includegraphics{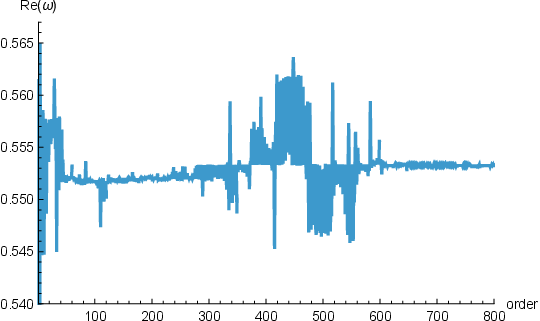}\includegraphics{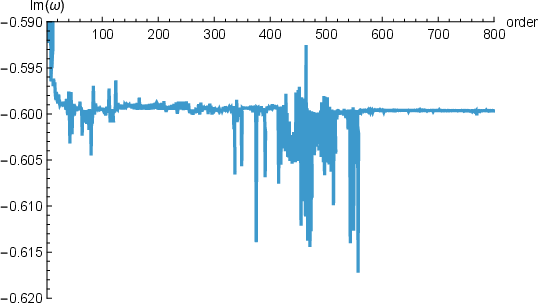}}
\caption{Example of slow convergence: $r_0=1$, $\epsilon=0$, $a_{1}=0.5$, $a_{2}=10$ ($\ell=n=1$). The Frobenius method yields the quasinormal frequency $\omega = 0.583998 - 0.648969 i$. By contrast, the WKB approximation gives $\omega = 0.552122 - 0.598878 i$ at 200th order, $\omega = 0.553334 - 0.599359 i$ at 400th order, $\omega = 0.552850 - 0.602733 i$ at 500th order, and $\omega = 0.553342 - 0.599663 i$ at 800th order. Thus, even by the 800th order, convergence to the accurate numerical result is not achieved.}\label{fig:A10}
\end{figure*}

\begin{figure*}
\resizebox{\linewidth}{!}{\includegraphics{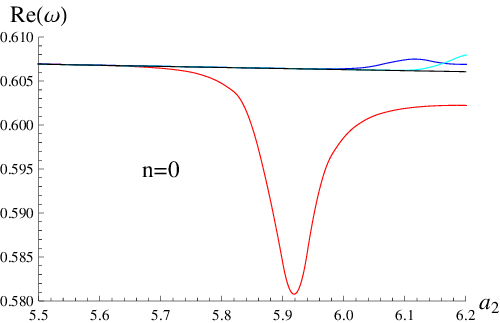}\includegraphics{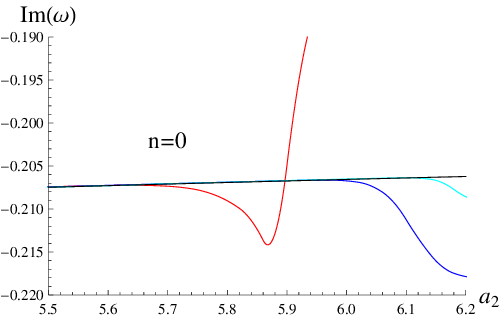}}
\resizebox{\linewidth}{!}{\includegraphics{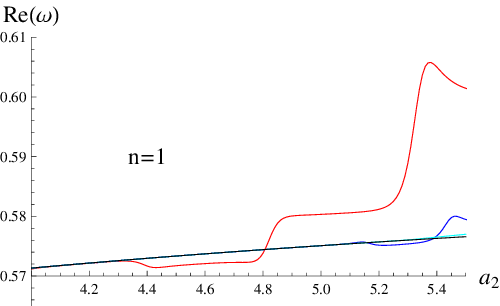}\includegraphics{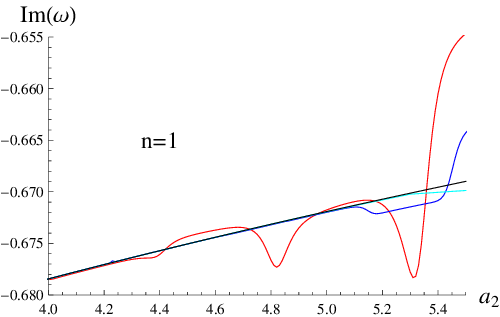}}
\caption{Real and imaginary parts of the test scalar field quasinormal frequencies ($\ell = 1$, $n = 0,1$) in the background of a parametrized black hole ($r_{0}=1$, $\epsilon=0$, $a_{1}=0.5$) as functions of $a_{2}$. Results are computed using the diagonal Padé approximants at different WKB orders: 100th (red), 200th (blue), 300th (cyan), and 500th (black), the latter coinciding with accurate reference values. As $a_{2}$ increases, the configuration departs from the moderately behaved regime, and convergence becomes progressively slower.}\label{fig:convergence}
\end{figure*}

We reserve the term \emph{apparent convergence} for a different situation: over a long but finite interval of orders the sequence may cluster around a value $\omega_{\mathrm{app}}$ that is still far from $\omega_{\mathrm F}$. Figure~\ref{fig:A10} illustrates this behavior. There the diagonal Padé approximants oscillate around a wrong plateau, and the truly asymptotic regime is reached only at much higher orders, if it is reached at all within the displayed range. In practice this phenomenon is typical of non-moderate metrics, whose stronger near-horizon deformation delays the onset of reliable high-order stabilization~\cite{Konoplya:2022pbc}.

The transition between the convergent and apparently convergent regimes is illustrated in Fig.~\ref{fig:convergence} by comparing the diagonal Padé approximants at several fixed WKB orders: $N=100$, $200$, $300$, and $500$. As the truncation order increases, the critical value of $a_2$ at which the Padé result starts to deviate significantly from $\omega_{\mathrm F}$ shifts toward larger values with increasing order. This behavior indicates that part of the observed loss of accuracy at moderate orders is not a genuine breakdown of the asymptotic method, but rather a delayed onset of the asymptotic regime, which can be pushed deeper into the non-moderate sector by going to sufficiently high WKB orders. At the same time, the WKB order required to achieve a reasonable accuracy grows very rapidly as $a_2$ increases, indicating that strongly nonmoderate configurations enter the true asymptotic regime only at extremely high orders. For higher overtones this effect becomes substantially stronger: the critical value of $a_2$ decreases rapidly with the overtone number, while the number of WKB orders needed to obtain a reasonable approximation increases even faster.

\section{Conclusions}\label{sec:conclusions}

We have shown that very-high-order WKB, when combined with diagonal and near-diagonal Padé approximants, becomes a powerful practical tool for problems that are difficult for the standard low-order treatment, in particular the first overtones with $n>\ell$ and the long-lived quasinormal modes of massive fields. In these regimes the high-order sequence is usually not monotonic in the WKB order: the approximants oscillate, and only their overall stabilization at sufficiently high orders carries reliable information. Therefore the absence of order-by-order convergence does not by itself indicate a failure of the method.

At the same time, the high-order WKB output must be interpreted with care. In non-moderate metrics, especially when the near-horizon geometry is strongly deformed, the Padé sequence may cluster for many orders around a value which is still far from the accurate result, as illustrated in Fig.~\ref{fig:A10}. We have referred to this behavior as apparent convergence. Thus numerical stabilization alone is not always a sufficient criterion of correctness, and whenever possible it should be checked against an independent method such as the Frobenius approach.

The massive-field case is particularly interesting. Although the standard WKB derivation is based on a two-turning-point picture, the resummed high-order scheme often yields accurate quasinormal frequencies even when the effective potential develops a third turning point, provided that the dominant barrier maximum is preserved. This indicates that the high-order information encoded near the main peak remains highly useful beyond the strict textbook assumptions of the method.

Overall, very-high-order WKB should be viewed not as a naive order-by-order expansion, but as a resummed asymptotic scheme. In this form it provides an efficient and versatile method for overtones, quasi-resonances, and other otherwise difficult spectral problems, while its main practical limitation is the possibility of false stabilization at insufficiently high orders.

While alternative numerical approaches exist, such as the Leaver method~\cite{Leaver:1985ax}, which is based on a strictly convergent procedure and allows one to determine quasinormal frequencies with high precision, the WKB method possesses several practical advantages. In particular, it does not require rewriting the wave equation in a rational form and, unlike the Leaver method, does not rely on an initial guess for the frequency. The latter typically necessitates scanning the parameter space with very small steps in order to avoid jumping between different modes, which significantly increases the computational cost. In contrast, the WKB method can be applied directly across a wide range of parameters without the need for an initial guess, making it considerably more efficient.

\appendix
\section{Description of the code}

\begin{itemize}
\item \texttt{WKBOrdersList}: the main routine. It takes the value of the effective potential and its derivatives at the point of maximum, passes them to \texttt{BenderWu}, and returns the WKB corrections for a chosen overtone number.
\item \texttt{PrecisePotentialPeak}: locates the maximum of the effective potential. It searches for the relevant extremum in the interval between the event horizon and the cosmological horizon (or spatial infinity), determined from the lapse function, and returns the point used by \texttt{WKBOrdersList}.
\item \texttt{WKBOrdersPadefication}: constructs the Padé approximants from the list of WKB corrections produced by \texttt{WKBOrdersList}.
\item \texttt{WKBOrdersWynn}: applies the fast Wynn epsilon algorithm to the output of \texttt{WKBOrdersList} and produces the diagonal Padé approximants at even orders and the superdiagonal Padé approximants at odd orders, as described in the appendix of~\cite{Matyjasek:2019eeu}.
\item \texttt{WKBOrdersWynnDiagonal}: extracts only the diagonal Padé approximants.
\item \texttt{WKBOrdersEWynn}: applies the same Wynn epsilon algorithm to the sequence supplemented by the vanishing zeroth-order WKB term, thereby generating the subdiagonal Padé approximants.
\item \texttt{CesaroMean}: computes the Cesàro mean of the input sequence.
\end{itemize}

\bibliography{bibliography}
\end{document}